# Origin of superconductivity in tungsten thin films

Vivas Bagwe[*], Rishabh Duhan[†], Bhagyashree Chalke, Jayesh Parmar, Somak Basistha and Pratap Raychaudhuri[‡]

*Tata Institute of Fundamental Research, Homi Bhabha Road, Colaba, Mumbai 400005, India.*

The most common allotrope of tungsten, $\alpha-\mathrm{W}$, has a superconducting transition at a temperature of ~11 mK. However, two other forms of tungsten have been reported to have superconducting transitions in the temperature range $T_c$ ~ 2-5 K when synthesized as thin films: Crystalline $\beta-\mathrm{W}$ and amorphous W ($a-\mathrm{W}$). In this work we carry out a systematic study of W films synthesized using d.c. magnetron sputtering, using transport, low frequency magnetic shielding response, and transmission electron microscopy. Our results show that while $a-\mathrm{W}$ is indeed a bulk superconductor, superconductivity in $\beta-\mathrm{W}$ probably originates from an amorphous phase that forms along with the $\beta-\mathrm{W}$ phase. Our findings reconcile some of the anomalies earlier reported in $\beta-\mathrm{W}$, such as the very small superconducting gap and the decrease of $T_c$ with increase in film thickness.

[*] vivas@tifr.res.in
[†] rishabh.duhan@tifr.res.in
[‡] pratap@tifr.res.in

I. Introduction

Bulk tungsten crystallizes in body-centred-cubic $\alpha-W$ phase and has a very low superconducting transition temperature[1], $T_c \sim 11$ mK. However, tungsten thin films often show a much larger superconducting transition temperature[2,3] with $T_c \sim$ 2-5 K. Early reports attributed this enhanced superconducting transition to the metastable $\beta-W$ phase with A15 crystal structure[4]. $\beta-W$ also has a large spin-orbit coupling[5], which along with superconductivity makes it a potential candidate for unconventional superconductivity[6] and spintronic applications[7,8]. At the same time, amorphous W ($a-W$) thin films with $T_c$ as high as 5 K can also be grown by incorporating small amounts of impurities during the deposition process. One way to obtain $a-W$ films is by breaking down an organometallic precursor like W(CO)$_6$ using a focussed ion beam[9,10,11], where the main impurities are carbon, oxygen and gallium. Alternatively, $a-W$ can also be synthesised using magnetron sputtering, where small amounts of nitrogen or oxygen is introduced in the deposition process[12,13]. However, despite considerable work, the origin of superconductivity in Tungsten films has remained controversial[14]. Many reports suggest that superconductivity in $\beta-W$ is stabilised only in the presence of disorder, either through the formation of nanocrystalline structures, or in the presence of second impurity phase like[12,13,14], WN$_x$. It has been reported that in $\beta-W$ films, the superconducting transition temperature gets suppressed when the thickness is increased in the tens of nanometers range, which is in contrary to the behaviour observed in most superconducting films. Furthermore, early tunnelling studies in $\beta-W$ films using planar tunnel junctions revealed a large sample to sample variation in the superconducting energy gap ($\Delta$), with gap to $T_c$ ratio[3], $\frac{\Delta}{k_B T_c} \sim 1.1 - 1.8$ (where $k_B$ is the Boltzmann constant). More recent optical conductivity measurements found the ratio to be[15] $\frac{\Delta}{k_B T_c} \sim 1$, which is much smaller than the expected value of 1.76 from Bardeen-Cooper-Schrieffer (BCS) theory. While such a small ratio could in-principle arise from isolated Fermi surface pockets that are much weaker coupling than the rest of the Fermi surface, extrinsic effects such as a normal metal phase coexisting with the superconductor needs to be carefully ruled out. The situation gets further complicated since $\beta-W$ films often need to be grown in the presence of small amount of an impurity gas like N$_2$ to minimise the $\alpha-W$ phase that inevitably grows along with it. In fact, a recent work suggests that the superconductivity in $\beta-W$ films might be related to disordered phases that grow alongside the $\beta-W$ phase[13].

In this work, we investigate the origin of superconductivity in superconducting tungsten films grown through d.c. magnetron sputtering. Controlling the deposition parameters, we are able to grow $\beta - W$ and $a - W$ films. We study the properties of both $\beta - W$ and $a - W$ films using a combination of structural, electrical, magnetic measurements and low temperature scanning tunnelling spectroscopy (STS). We observe that $a - W$ behaves like a weak coupling BCS superconductor. In contrast superconductivity in the putative $\beta - W$ films is highly anomalous: While bulk measurements show a clear superconducting transition, we do not observe a superconducting energy gap when the film is probed using STS. Cross sectional TEM studies indicate that the superconductivity might not be associated with the $\beta - W$ phase but with an amorphous W layer that forms underneath the $\beta - W$ phase. The central result of this paper is that contrary to popular belief crystalline $\beta - W$ might not be a superconductor.

## II.  Experimental details

W films were grown from a 2 inch diameter W target using d.c. magnetron sputtering on oxidized Si, sapphire and Nb-doped STO substrates at room temperatures in Ar/N$_2$ gas mixture. In this work we will focus on films grown on 0.5 mm thick Si substrates with approximately 200 nm of thermal oxide on the top. The base pressure of the chamber prior to deposition was $8 \times 10^{-7}$ mbar while the depositions were carried out at 10 mbar pressure. The composition of the gas mixture during growth was controlled with mass flow controllers. The flow rate of Ar was fixed at 27 sccm whereas N$_2$ flow rate was varied between 0-1 sccm. The target substrate distance was fixed at 4.5 cm and the sputtering power was 60 W.

For all films structural characterisation was performed using X-ray diffraction (XRD). $\theta$-$2\theta$ XRD scans were acquired keeping an offset of $4^0$ with the plane of the film to avoid contribution from the substrate. Some films were also characterised using grazing incidence diffraction (GID) to get an insight of the structure at different thickness below the film surface. Transmission electron microscopy is performed using an aberration corrected (Cs corrected) FEI-TITAN microscope operated at 300 kV and equipped with a field-emission gun source. To prepare the TEM specimen, the film was cut in 1.7mm x 1mm dimension using a diamond micro-saw. A sandwich of two such samples was prepared using GATAN G1 epoxy and embedded in Ti slot grid of dimension 1.8 mm x 1 mm. The disk thus prepared was grinded to 80 micron thickness and then a dimple was made using dimple grinder leaving 10 micron thickness. This sample was then milled in a precision ion milling system using Ar ions. The sample was milled using top and bottom guns at 5kV, 5 deg angle till perforation in the sample was observed; after that the sample was further polished with low energy beam at 2kV and 1 kV. The sample stage was cooled with liquid $N_2$ in order to avoid the sample getting altered due to local heating.

Measurement of bulk superconducting properties were performed in a $^4$He cryostat fitted with a 100 kOe superconducting solenoid. Electrical transport measurement was performed using the conventional 4-probe method. Resistivity and Hall measurements were performed on samples deposited in a 6-probe Hall bar geometry using a shadow mask. The magnetic shielding response of the film was measured using a two-coil technique[16] where the film is sandwiched between a quadrupolar primary coil and a dipolar secondary coil and the in-phase and out-of-phase component of the mutual inductance ($M'$ and $M''$) is measured using a lock-in amplifier. In this setup the onset of superconductivity is marked by a rapid decrease in $M'$ (and the appearance of a dissipation peak in $M''$) since the superconducting film partially shields the magnetic field generated in the primary coil from the secondary, thus providing a non-contact method for determining $T_c$. STS measurements were performed using a home-built scanning tunnelling microscope[17] (STM) operating down to 410 mK and fitted with a 90 kOe superconducting solenoid. For STS measurements the sample was transferred in-situ from the chamber to the STM using an ultra-high-vacuum suitcase without exposure to air. This ensured a pristine un-oxidised surface for STS measurements.

The thickness of the films was measured using a stylus profilometer.

### III. Results

Over the course of this study more than 100 films were deposited by varying the thickness and the Ar/$N_2$ gas ratio. Fig. 1(a) shows representative θ–2θ scans for a set of W films grown with different flow rate of $N_2$ during the deposition process. All films here have thickness of $140 \pm 10$ nm. We observe the following general trend. For the films grown between 0 - 0.1625 sccm we observe pronounced peaks close to $35.5^0$ and $75.2^0$ corresponding to $[200]_\beta$ and $[400]_\beta$ peaks respectively. There is an additional peak at $72.6^0$. This peak most likely corresponds to the $[211]_\alpha$ peak from a strained $\alpha - W$ impurity phase. While there are some peaks of $W_2N_3$ in this range, this is unlikely to be its origin for two reasons: First, we do not observe the more intense peaks corresponding to the same structure and, second, the peak intensity does not increase with $N_2$ flow rate. In contrary, when $N_2$ flow rate is increased beyond 0.175 sccm, this peak disappears completely. Furthermore, the presence of $\alpha - W$ impurity in the films can also be seen from the position of the peak close to $40^0$ which is an admixture of of the $[110]_\alpha$ peak (at $40.26^0$) and $[210]_\beta$ peak (at $39.88^0$). For the sample grown in pure Ar atmosphere, this peak is in between the expected value for $[110]_\alpha$ and $[210]_\beta$ showing that there is significant contribution from both phases. However, as the N$_2$ flow rate is increased from 0 to 0.1625 sccm we observe that this peak shifts towards the expected value of $\beta - W$ phase showing a gradual stabilisation towards the $\beta - W$ phase. The same can be seen from the intensity ratio of $[211]_\alpha$ and $[400]_\beta$ peaks which decreases from $\frac{I_{[211]\alpha}}{I_{[400]\beta}} \sim 1$ to $\frac{I_{[211]\alpha}}{I_{[400]\beta}} \sim 0.2$ in the same range (Fig. 1(c)). However, with further increase of N$_2$ in the sputtering gas the film gets completely amorphised. Films grown with N$_2$ flow rate between 0.175-1 sccm only show very broad humps close to $35^0$ and $65^0$ and no crystalline peak (Fig. 1 (b)) consistent with the formation of $a - W$.

Of the films deposited, we obtained superconductivity most consistently in the $a - W$ films with the superconducting transition varying in the range, $T_c \sim 4.4 - 5.2$ K. Fig. 2(a) shows the representative magnetic shielding response of one $a - W$ film (140 nm thickness) with $T_c$ ~ 4.8 K. The single sharp drop in $M'$ at $T_c$ and the clean dissipative peak in $M''$ are both signatures of clean homogeneous film. Fig. 2(b) show the resistance ($R$) versus temperature ($T$) at various magnetic field. From the $R$-$T$ we extract in Fig. 2(c) the temperature variation of the upper critical field, $H_{c2}$, where the resistance reaches 95% of the normal state value. To further understand the superconducting state, we investigated a similar sample using STM. Fig. 3(a) shows the topography of the sample. The root mean square surface roughness is ~0.7 nm. Fig. 3(b) shows the normalised differential conductance $\left(G_N(V) = \frac{dI}{dV}\Big|_V / \frac{dI}{dV}\Big|_{4\ mV}\right)$ as a function of voltage at different temperatures, averaged over a uniformly placed $8 \times 8$ grid over this area. The $G_N(V) - V$ curves display symmetric coherence peaks and minima at low bias consistent with what is expected for a conventional superconductor. We fit each curve with the standard tunnelling equation[18], using the broadened Bardeen-Cooper-Schrieffer (BCS) density of states: $N(E) = Re\left\{\frac{|E+i\Gamma|}{\sqrt{(E+i\Gamma)^2-\Delta^2}}\right\}$, where $\Delta$ is the superconducting energy gap and $\Gamma$ is an additional broadening parameter that accounts for non-thermal sources of broadening[19]. The temperature variation of $\Delta$ (Fig. 3(c)) follows the expected BCS variation within experimental accuracy. However, $\frac{\Delta(0)}{k_B T_C} \sim 2.6$ is larger than the weak coupling BCS value of 1.76 suggesting that it is a strong coupling superconductor. Fig. 3(d) shows the conductance map acquired at bias voltage of $V = 1.5\ mV$ close to the coherence peak $\left(\frac{dI}{dV}\Big|_{V=1.5\ mV}\right)$, over the same area at 0.41 K with a magnetic field of 5 kOe applied perpendicular to the film plane. In the conductance map each vortex appears as a dark spots, corresponding to a local conductance minimum[20,21]. This identification was further confirmed by counting the number of vortices which is consistent with the total magnetic flux passing through this area.

Superconductivity in $\beta - W$ films were much less consistent than in amorphous films. Of the 22 $\beta - W$ samples that we measured 8 showed superconducting transition down to 2.3 K. From XRD we observed that all samples had some amount of $\alpha - W$ impurity, but we did not observe any systematic variation between the intensity of these impurity peaks and $T_c$ of the film. However, the $T_c$ of $\beta - W$ films were consistently lower than 4 K. Fig. 4(a) shows the R-T in in different magnetic fields for one such sample (thickness 140 nm) with $T_c \sim 3.85$ K. The zero-field transition is sharp and $H_{c2}$ (Fig. 4(b)) is of the same order of magnitude as $\alpha - W$. The normal state resistivity in about 20% lower that the amorphous film of the same thickness. The XRD for this film (Fig. 4(c)) shows clear peaks corresponding to the $\beta - W$ phase and very small contribution from $\alpha - W$ impurity.

Since $\beta - W$ films did not consistently show superconductivity, we performed STS measurements on several samples adopting the following protocol: First performed STS measurement on an uncontaminated surface without exposing to air; after completing STS measurements the sample was taken out and the $T_c$ was determined through magnetic shielding response measurements. While some of the samples showed a superconducting transition in the magnetic shielding response, none of the samples showed clear signature of superconductivity in STS measurements. Fig. 5(a) shows the topography of a $\beta - W$ film with $T_c \sim 3.8$ K. $G(V)$-$V$ curves were acquired over this area on a $16 \times 16$ grid at 410 mK in zero magnetic field. Fig. 5(b) shows the normalised zero bias conductance map over this area. The individual spectra either showed a broad V shape (Fig. 5(c)-(e)) extending up to high bias, or were nearly flat, and none of them showed any clear signature of superconductivity related features. However, a weak signature of superconductivity below 1 mV was observed when we inspected the average of all 256 spectra (Fig. 5(f)). Here in addition to the broad V shape and a small additional decrease in conductance (~ 5%) becomes visible below 1 mV. This weak low bias feature is reminiscent of proximity induced superconductivity that is observed when a superconductor is covered with a normal metal[22]. Fig. 5(g) shows the magnetic shielding response measured on the same film. The shielding response shows a drop in $M'$ at the superconducting transition associated with a dissipative peak in $M''$.

To understand the anomalous behaviour of $\beta-W$ films, we carried out detailed cross-sectional TEM (cs-TEM) measurements on the same sample. Figure 6(a)-(b) show the cs-TEM images acquired at the top and bottom (close to the substrate) of the film along with the selective area electron diffraction (SAED) pattern from the top and the bottom (Fig. 6(c)-(d)) taken with a 5 μm aperture. Fig. 6(e) and 6(f) show the expanded view of the areas in $10\ nm \times 10\ nm$ white box in Fig. 6(a) and 6(b) respectively. We observe that structure of the film is not the same at the top and the bottom. At the top of the film, we can clearly see crystalline structure in the image. The SAED pattern (Fig. 6(c)) contains large number of sharp spots forming a ring pattern which is expected when the diffraction is from several crystallites within the field view. In Fig.6(g) we show the 2-dimensional Fourier transform (2D-FT) Fig 6 (e); the 2D-FT of Fig. 6(e) shows sharp spots that we can index with the $\beta-W$ crystal structure. In contrast, no crystalline structure is discernible close to the substrate; here both the SAED (Fig. 6(d)) and the 2D-FT of the area shown in Fig. 6(f) (Fig. 6(h)) show diffuse rings corresponding to either a nanocrystalline or an amorphous state.

To understand the evolution of the structure more carefully, we recorded nanobeam diffraction (NBD) patterns with electron beam spot size of 1.5 nm (Fig. 7(b)), at 23 locations across the cross-section of a 220 nm thick film, where location 1 is close to the top of the film and location 24 is close to the substrate (Fig. 7(a)). The panels labelled 1-23 show representative NBD patterns at 14 locations. The NBD pattern at location 1 shows sharp Bragg spots corresponding to a crystalline structure. These Bragg spots are visible till location 4 which is about 29 nm from the film surface. Beyond this the NBD pattern displays only a ring that progressively becomes diffuse as we go towards the substrate. This implies the presence of either of nano-crystallites or an amorphous structure with short range correlations. Beyond location 15 (corresponding to 133 nm from the film surface), the ring also disappears, and we see only a diffuse intensity corresponding to a completely amorphous structure. These results strongly suggest the existence of approximately 90 nm thick amorphous W beneath the $\beta-W$ film. In contrast, similar measurements performed on $a-W$ film does not show any structural variation within the film thickness and show an amorphous structure everywhere.

To further corroborate this conclusion, we deposited a series of films at 0.16 sccm $N_2$ flow during deposition with varying thickness (by varying the time of deposition) and performed XRD of each of the films. Fig. 8 shows the θ–2θ scans for this set of films. Up to a thickness of 35 nm we do not observe any Bragg peak showing that the film remains amorphous. At 53 nm broad humps start appearing around the positions where Bragg peaks are expected for the $\beta-W$ structure, indicating the formation of nanocrystals. Clear Bragg peaks corresponding to $\beta-W$ appear at 70 nm thickness. Between 70 nm to 140 nm the Bragg peaks get more intense. For the 140 nm thick film we can identify the $[211]_\alpha$ peak showing the presence of $\alpha-W$ impurity, consistent with our earlier discussion on films with similar thickness. The inset shows the variation of $T_c$ for films of different thickness deposited under the same conditions. Consistent with the structural observations, the 14 nm amorphous film has $T_c \sim 4.65$ K. $T_c$ decreases with film thickness and for thickness >50 nm where the $\beta-W$ appears, $T_c < 4$ K.

### IV. Discussion

We can now put these results in perspective. We have consistently observed superconductivity only in $a-W$ films grown in the presence of moderate N$_2$ impurities during growth. Since we cannot identify any crystalline WN$_x$ phase, we conclude that the N$_2$ probably gets incorporated as interstitial atoms, thereby destroying the crystal structure and rendering the film amorphous. In contrast, the origin of superconductivity in films that show pronounced $\beta-W$ peaks in XRD is more complex. First, the superconducting transition temperature of these films vary widely, and many films turn out to be non-superconducting down to 2.3 K. Those which exhibited superconductivity had transition temperature < 4 K, which is smaller than $a-W$ films. However, even for films that showed a superconducting transition from transport and magnetic measurements, we observed only a very weak signature of proximity induced superconductivity at the surface of the film using STM. Combining the information obtained from structural measurements, we conclude that superconductivity originates from the amorphous tungsten underlayer, whereas $\beta-W$ itself remains non-superconducting. The thickness of this amorphous underlayer is expected to depend strongly on the nature of the substrate. Since many different Si wafers were used in this study, the thickness of the thermal oxide as well as its local structure could vary from one substrate to another. This probably is one of the reasons for the large sample to sample variation in the $T_c$ of $\beta-W$ films with notionally the same thickness. Another reason for this variation is that the complete amorphization of tungsten films happens within a narrow window of N$_2$ flow-rate during growth. While the $\beta-W$ phase with least amount of $\alpha-W$ impurity is obtained for a N$_2$ flow rate of 0.1625 sccm, a slight increase to ~0.175 sccm results in a complete amorphization of the films into $a-W$. This sensitivity makes it difficult to reproduce the same deposition condition in every run, resulting in variations in the thickness of the amorphous underlayer vis-à-vis the crystalline $\beta - W$ layer. We would like to note that few $\beta - W$ films that we grew of Nb-doped SrTiO3 and sapphire also did not show any superconductivity.

In the light of our results, we can now reexamine the unusual superconducting properties of $\beta - W$ films. In general, in the presence of sharp diffraction peaks from $\beta - W$, it is easy to miss the presence of the amorphous layer from XRD measurements. However, we note that if superconductivity arises from the amorphous underlayer, the normal $\beta - W$ film above will influence superconductivity in two different ways. First, weak superconductivity will get induced through proximity effect in the intrinsically normal metal $\beta - W$ thus explaining the proximity induced minigap[22] in our experiments. On the other hand, the $T_c$ of the amorphous superconducting underlayer would also get partially suppressed due to inverse proximity effect, explaining why films showing the $\beta - W$ phase always show $T_c$ that is smaller than pure $a - W$ films. This observation is consistent with an earlier report where a 60 nm thick $\beta - W$ film was seen to have lower $T_c$ compared to a 35 nm film[14]. It is pertinent to note that in ref. 14 the 35 nm film only showed a very broad peak in XRD suggesting an amorphous structure as opposed to the sharp diffraction peaks observed in the 60 nm film, which is again consistent with our results.

Finally, we would also like to comment on the optical conductivity measurements[15] that reveal an anomalously small superconducting energy gap in $\beta - W$. While a gap ratio, $\frac{\Delta}{k_B T_c}$, larger than the expected BCS value of 1.76 is often observed in strong coupling superconductors[23], a ratio that is smaller than the BCS ratio is rare. In principle, such a feature can be explained from the intrinsic anisotropic superconducting gap where parts of the Fermi surface have small gap or from the presence of metallic regions in the film which dissipate energy at high frequencies. In the light of our results, we believe that the small gap is related to the superconducting/non-superconducting sandwich structure in $\beta - W$ films. We note that unlike STM which probes the samples surface, THz optical conductivity performed in the transmission geometry probes the entire depth of the film which will give an average response from both the superconducting amorphous and non-superconducting $\beta - W$ layers. This would naturally explain the smaller than expected depletion of the real part of the conductivity below $T_c$ and a small $\Delta$, something that should be quantitatively verified from detailed calculations.

**V.    Conclusion**

In this paper, we revisited the decades old problem on the origin of superconductivity in Tungsten thin films. Our results show that $a-\text{W}$ is a conventional strong coupling superconductor. On the other hand, contrary to common belief, crystalline $\beta-\text{W}$ films might not be a superconductor at all. Here, superconductivity originates from an amorphous tungsten phase that form along with the $\beta-\text{W}$ phase. A similar conclusion was recently arrived at in ref. 13 which focussed on primarily on X-ray photoemission measurements. In our study, we find that this amorphous phase forms as an underlayer below the $\beta-\text{W}$ phase. However, we would like to note that the precise microstructure of this layer could depend on the deposition condition and the substrate. Nevertheless, the non-intrinsic nature of superconductivity in $\beta-\text{W}$ provides plausible explanation of several unusual superconducting properties reported in this material and should be carefully considered in any future study of this material.

Acknowledgements: We thank Nilesh Kulkarni and Vilas Mhatre for their help with XRD measurements. This work was supported by the Department of Atomic Energy, Government of India. VB and RD deposited the tungsten thin films and coordinated the project. Structural characterisation using XRD and analysis of the data was primarily performed by VB. Low temperature transport and scanning tunnelling microscopy measurements and analysis were performed by RD. SB performed the two-coil mutual inductance measurement. TEM samples were prepared by JP and microscopy was done by BC. PR conceived the problem and supervised the project. The paper was written by PR with inputs from all authors.

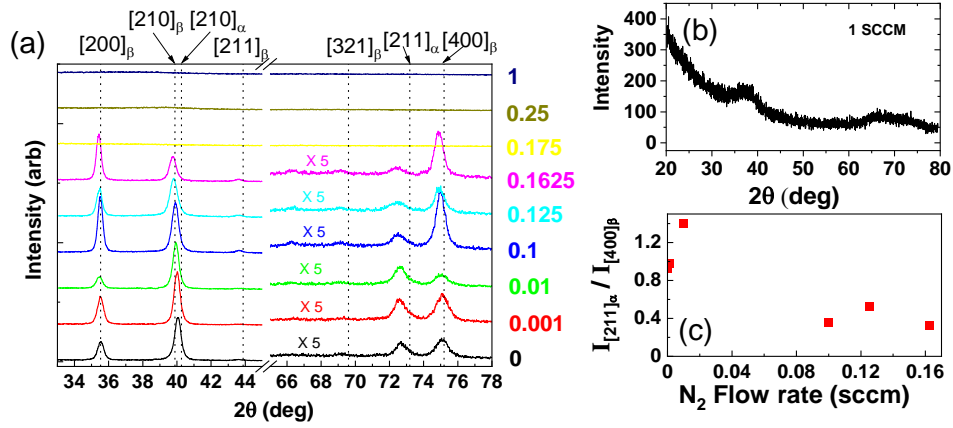

**Figure 1.** (a) XRD $\theta$–$2\theta$ scans for tungsten films grown in $N_2$/Ar atmosphere; the Ar flow-rate was fixed at 27 sccm whereas the $N_2$ flow-rates for different films are shown in the right of the panel (in units of sccm). Successive plots have been shifted upward for clarity. (b) XRD $\theta$–$2\theta$ scan of the tungsten films grown with 1 sccm flow rate of $N_2$. (c) Intensity ratio of $[211]_\alpha$ and $[400]_\beta$ peaks, $\frac{I_{[211]_\alpha}}{I_{[400]_\beta}}$, for films grown with different flow rates of $N_2$.

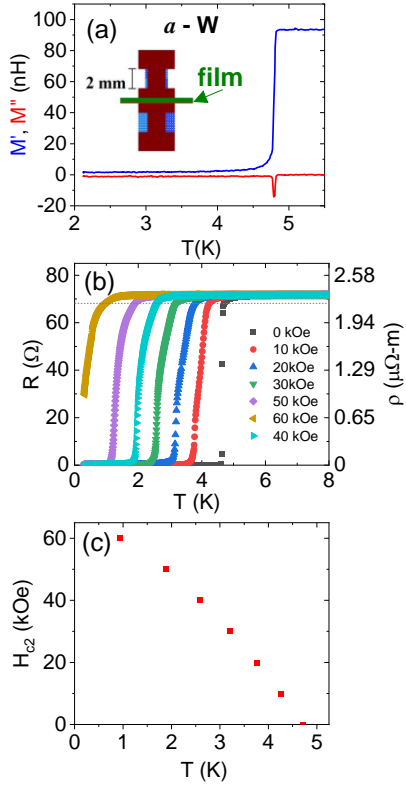

**Figure 2.** (a) Magnetic shielding response of a 140 nm thick $a-W$ film measured from the real ($M'$) and imaginary ($M''$) part of the mutual inductance. (*inset*) Schematics of the mutual inductance setup used to measure the magnetic shielding response of superconducting films. (b) Resistance/resistivity versus temperature of an $a-W$ film in different magnetic field. The dashed horizontal line shows 95% of the normal state resistance from which $H_{c2}$ is determined. (c) Temperature variation of $H_{c2}$.

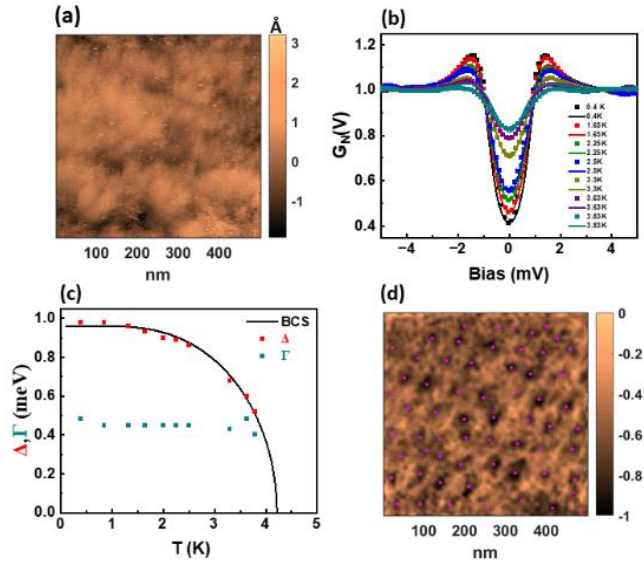

**Figure 3.** (a) Topographic map over $500\ nm \times 500\ nm$ area for a 140 nm thick $a-W$ thin film measured using STM at 4 K; the bias voltage is 15 mV and tunnelling current is 150 pA. (b) Normalised conductance $G_N(V)$ as a function of bias voltage at different temperature in zero magnetic field, averaged over 64 equally spaced points over this area; the conductance is normalised at 4 mV. The solid lines show the fits to the spectra. (c) Temperature variation of the superconducting energy gap, $\Delta$, and broadening parameter, $\Gamma$, extracted from the fits; the solid line is the expected BCS variation of $\Delta$. (d) Normalised $\frac{dI}{dV}\Big|_{V=1.5\ mV}$ map over the same area at 410 mK in a magnetic field of 5 kOe; the colour scale is offset subtracted and normalised is such that 0 and 1 correspond to the minimum and maximum conductance values respectively. The vortices appear at local minima in the conductance map and are shown as purple points.

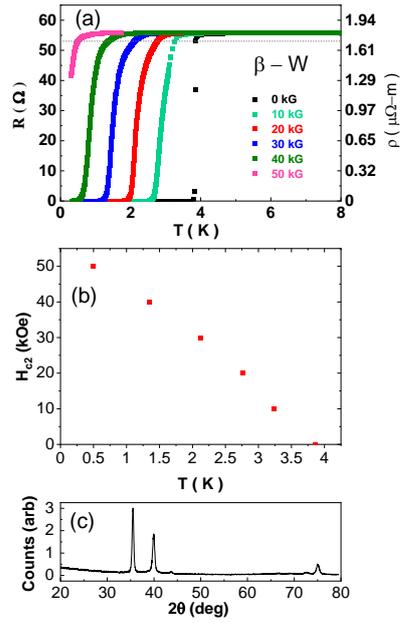

**Figure 4.** (a) Resistance/resistivity versus temperature of a $\beta-W$ film in different magnetic field. The dashed horizontal line shows 95% of the normal state resistance from which $H_{c2}$ is determined. (b) Temperature variation of $H_{c2}$. (c) XRD $\theta$–$2\theta$ scan showing the diffraction peaks corresponding to $\beta-W$.

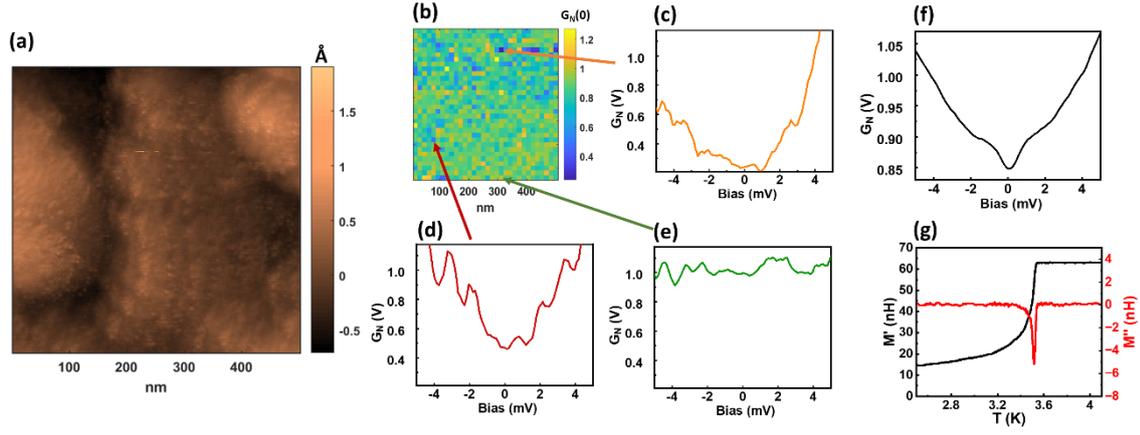

**Figure 5.** (a) Topographic map over $500\ nm \times 500\ nm$ area for a 140 nm thick $\beta - W$ thin film measured using STM at 4 K; the bias voltage is 15 mV and tunnelling current is 150 pA. (b) Normalised zero bias conductance $\left(G_N(0) = \dfrac{\left.\dfrac{dI}{dV}\right|_{V=0}}{\left.\dfrac{dI}{dV}\right|_{V=4\ mV}}\right)$ map over the same area. (c)-(e) Representative $G_N(V)\ vs.\ V$ spectra at three different locations. (f) Average Normalised conductance $G_N(V)$ vs. $V$ spectra over the entire area; a small suppression in conductance is observed below 1 mV riding over the overall V shape extending to high bias. (g) Magnetic shielding response for the same film measured from the real ($M'$) and imaginary ($M''$) part of the mutual inductance.

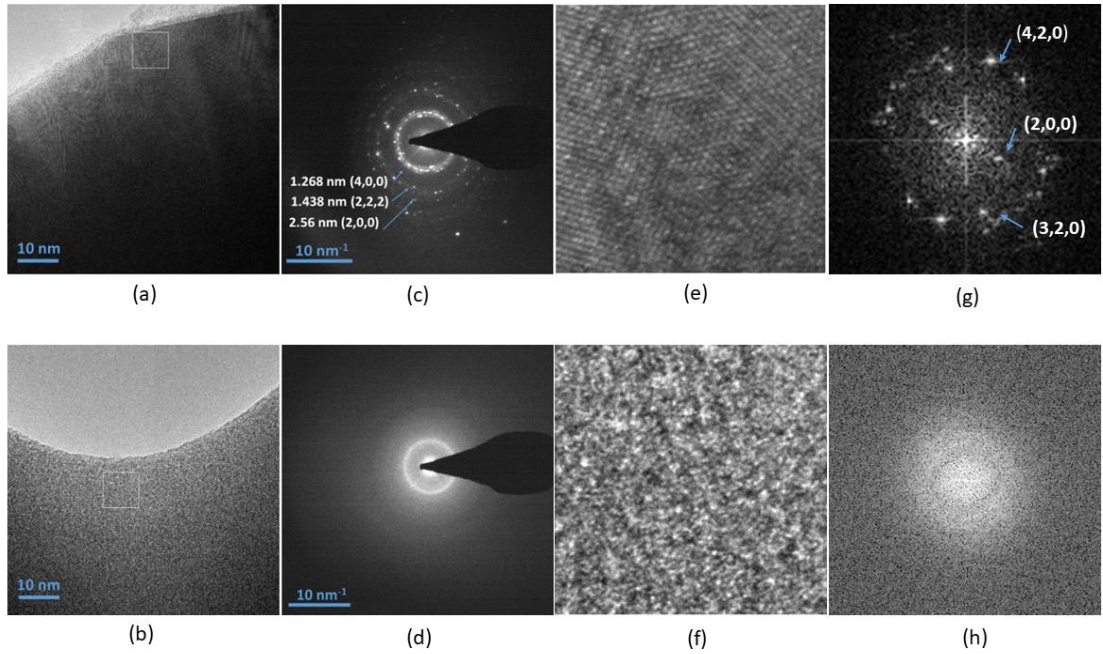

**Figure 6.** (a)-(b) Transmission electron micrograph of the top (a) and bottom (b) of a 220 nm thick $\beta - W$ thin film. (c)-(d) SAED pattern from the top (c) and bottom (d) of the film. (e)-(f) Expanded view of the of the $10\ nm \times 10\ nm$ area within the white box in (a) and (b) respectively. (g)-(h) 2D-FT of images (e) and (f) respectively; some Bragg spots in (g) are indexed with the $\beta - W$ structure.

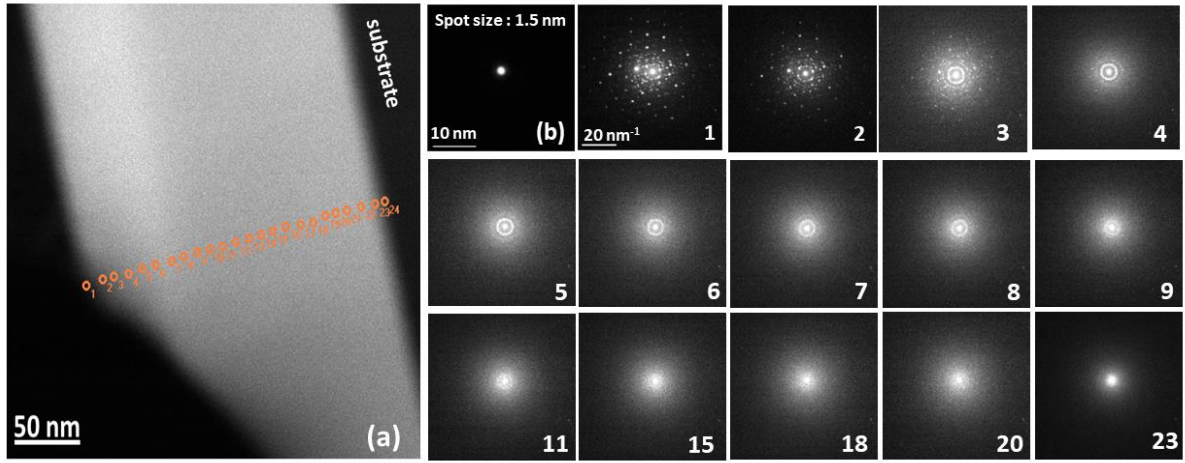

**Figure 7.** (a) Cross sectional view of a $\beta - W$ film. The point marked 1 is at the top of the film, while 24 is at the bottom. (b) Image of the 1.5 nm nanobeam used for the NBD acquired at locations 1-24. The panels marked 1-23 show the NBD patterns at the location marked by the same number on panel (a). We observe a gradual change from crystalline to amorphous structure as one goes from the top to the bottom of the film.

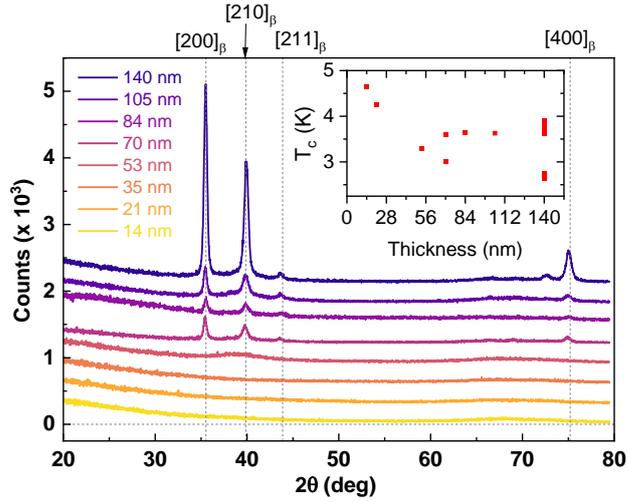

**Figure 8.** XRD $\theta$–$2\theta$ scans of tungsten films of different thickness grown in 0.16 sccm $N_2$ and 27 sccm Ar flow. Films with thickness up to 53 nm are amorphous. Peaks corresponding to $\beta - W$ start gradually appearing for films with thickness larger than 70 nm and become more intense with increase in thickness. Each successive graph has been shifted upward for clarity. (*inset*) Variation of $T_c$ with film thickness grown under same deposition conditions; for some thickness we have shown $T_c$ for multiple films of the same thickness to give an idea of sample-to-sample variation, which becomes more pronounced in the thickness range in which we observe $\beta - W$ crystalline peaks.